\documentclass[11pt]{article}

\usepackage[preprint]{acl}

\usepackage{times}
\usepackage{latexsym}
\usepackage{enumitem}
\usepackage[T1]{fontenc}

\usepackage[utf8]{inputenc}

\usepackage{microtype}
\usepackage{algorithm}
\usepackage{booktabs}
\usepackage{multirow} 
\usepackage{placeins}
\usepackage[table]{xcolor}
\usepackage{inconsolata}
\usepackage{amsmath}
\usepackage{graphicx}      
\usepackage{subcaption}    
\usepackage{graphicx}
\usepackage{amssymb}
\usepackage{xcolor,colortbl}
\usepackage{xcolor}
\usepackage{dblfloatfix}
\usepackage{tcolorbox}
\tcbuselibrary{skins, breakable}

\definecolor{c3}{cmyk}{0.3081,0,0.7209,0.3255}

\newtcbox{\hlprimarytab}{on line, rounded corners, box align=base,
  colback=c3!10,colframe=white,size=fbox,arc=3pt,
  before upper=\strut, top=-2pt, bottom=-4pt,
  left=-2pt, right=-2pt, boxrule=0pt}

\newtcbox{\hlsecondarytab}{on line, box align=base,
  colback=red!10,colframe=white,size=fbox,arc=3pt,
  before upper=\strut, top=-2pt, bottom=-4pt,
  left=-2pt, right=-2pt, boxrule=0pt}
\newtcolorbox{SummaryBox}{
  enhanced,
  breakable,
  colback=blue!5,        
  colframe=blue!40,      
  boxrule=0pt,           
  leftrule=3pt,          
  rightrule=0pt,
  toprule=0pt,
  bottomrule=0pt,
  arc=3pt,               
  left=1pt,
  right=1pt,
  top=1pt,
  bottom=1pt,
}

\newcommand{\dashifted}{\raisebox{0.5\depth}{\tiny$\downarrow$}}

\newcommand{\da}[1]{{\scriptsize\hlsecondarytab{\dashifted#1}}}

\usepackage{tcolorbox}
\tcbuselibrary{listings,breakable}
\usepackage{changepage}

\usepackage{listings}
\usepackage{xcolor}

\definecolor{promptbg}{RGB}{246,246,246}
\definecolor{promptframe}{RGB}{180,180,180}

\lstdefinestyle{promptstyle}{
    basicstyle=\ttfamily\small,
    backgroundcolor=\color{promptbg},
    frame=none,
    breaklines=true,
    breakatwhitespace=true,
    columns=fullflexible,
    keepspaces=true,
    showstringspaces=false
}
\newcommand{\ourmodel}{\textsc{CodeMEM}}
\title{
  \ourmodel{}: AST-Guided Adaptive Memory for Repository-Level Iterative Code Generation
}

\author{
Peiding Wang$^{1}$,
Li Zhang$^{1}$,
Fang Liu\thanks{Corresponding author.}$^{1}$,
Chongyang Tao$^{1}$,
Yinghao Zhu$^{2}$ \\
$^1$School of Computer Science and Engineering,
Beihang University \\
$^2$School of Computing and Data Science, The University of Hong Kong\\
{\tt \{wangpeiding, fangliu\}@buaa.edu.cn}
}

\begin{document}
\maketitle

\begin{abstract}
Large language models (LLMs) substantially enhance developer productivity in repository-level code generation through interactive collaboration. However, as interactions progress, repository context must be continuously preserved and updated to integrate newly validated information. Meanwhile, the expanding session history increases cognitive burden, often leading to forgetting and the reintroduction of previously resolved errors. Existing memory management approaches show promise but remain limited by natural language-centric representations.
To overcome these limitations, we propose \ourmodel{}, an AST-guided dynamic memory management system tailored for repository-level iterative code generation. Specifically, \ourmodel{} introduces the \textit{Code Context Memory} component that dynamically maintains and updates repository context through AST-guided LLM operations, along with the \textit{Code Session Memory} that constructs a code-centric representation of interaction history and explicitly detects and mitigates forgetting through AST-based analysis.
Experimental results on the instruction-following benchmark CodeIF-Bench and the code generation benchmark CoderEval demonstrate that \ourmodel{} achieves state-of-the-art performance, improving instruction following by 12.2\% for the current turn and 11.5\% for the session level, and reducing interaction rounds by 2–3, while maintaining competitive inference latency and token efficiency\footnote{The code and data are available at \url{https://github.com/zhu-zhu-ding/CodeMem}}.

\end{abstract}

\section{Introduction}

\begin{figure}[t]
    \centering
    \setlength{\abovecaptionskip}{0.1cm}
    \includegraphics[width=\linewidth]{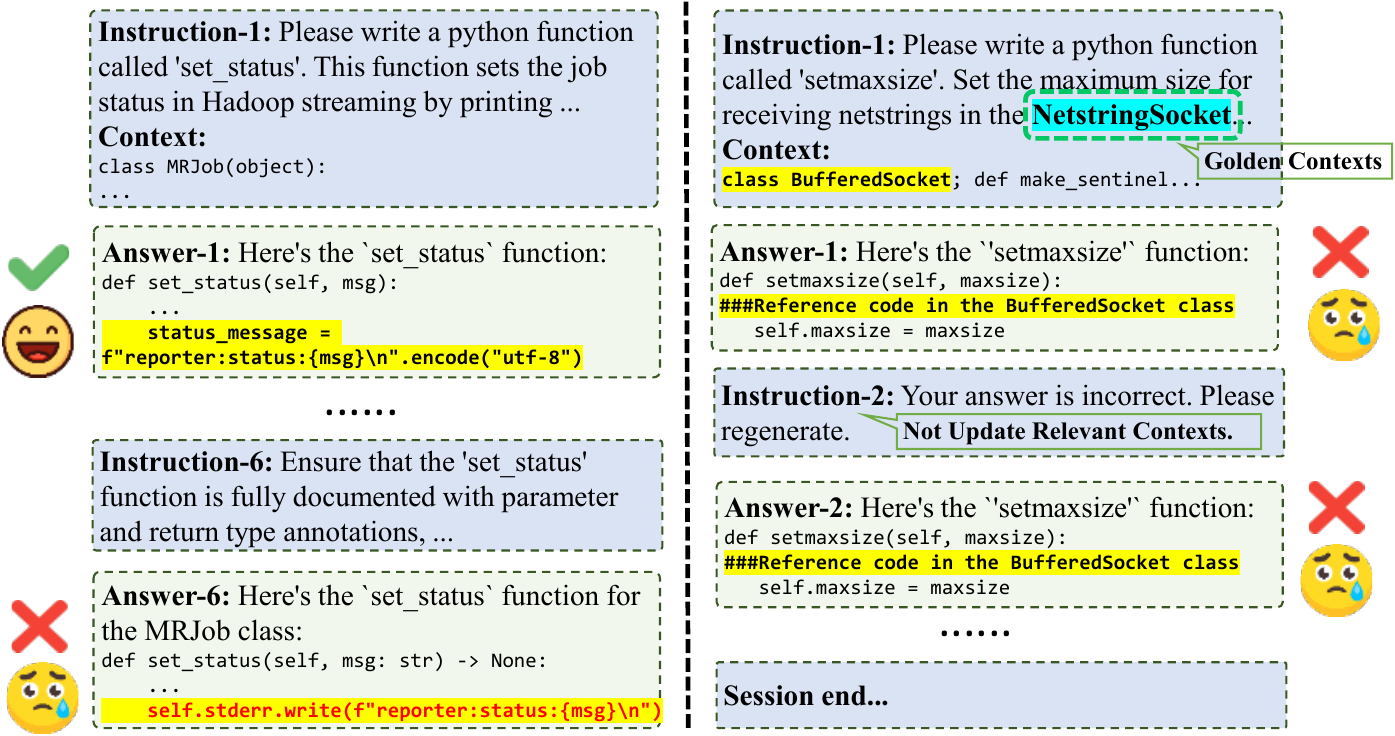}
    \caption{Examples of failure cases in iterative code generation. Left: As interactions progress, the LLM exhibits forgetting and overwrites previously correct modifications. Right: Static context handling propagates erroneous information and prevents the incorporation of relevant contexts, resulting in repeated incorrect code.}
	\label{fig:intro}
    \vspace{-0.5cm}
\end{figure}

In recent years, Large Language Models (LLMs)~\cite{deepseek,qwen25coder} in code generation has substantially improved developers’ productivity, especially in repository-level code generation tasks, where awareness of rich intra-repository context is essential~\cite{rl-coder,aixcoder7bv2}.
However, real-world development workflows commonly involve multi-turn interactions, such as refining requirements or requesting fixes for erroneous code~\cite{codeifbench,sreval}. We refer to this as \textbf{repository-level iterative code generation}.
As illustrated in Figure~\ref{fig:intro}, the code context in such scenarios is inherently dynamic. Satisfying evolving session instructions requires selectively preserving critical historical code context while continuously integrating newly relevant repository information. As interactions span multiple rounds, increasingly long and complex session histories impose growing cognitive burdens on LLMs, leading to forgetting, overwriting, or contradicting previously correct code modifications~\cite{lostmultiturnconversation,longcodearenaset}.

Memory management systems~\cite{memorybank,memagent,memos,memgpt} offer a promising direction for supporting long-term interactions by augmenting LLMs with external memory. However, existing approaches are primarily designed for natural language generation and rely on natural-language-centric memory representations. When applied to repository-level iterative code generation, such representations often blur code structure and behavior, hindering the dynamic retention of effective historical code context and the integration of newly relevant repository information. Moreover, session memory based on LLM summaries or dialogue retrieval fails to capture the evolving interaction state, leading to rigid behaviors where LLMs may forget prior corrections or repeatedly reintroduce resolved errors~\cite{lostmultiturnconversation,mt-bench}.

To address these limitations, we propose \ourmodel{}, a memory management mechanism specifically designed for repository-level iterative code generation. \ourmodel{} consists of two memory components:

\noindent\textbf{1) Code Context Memory} dynamically preserves effective historical code context while integrating newly relevant repository information to support evolving session instructions. \ourmodel{} uses an AST-guided selection mechanism to retain useful history and an LLM to decide when updates are needed. Upon updating, newly retrieved code is incrementally merged with retained context, allowing the memory to remain effective with the session.

\noindent\textbf{2) Code Session Memory} addresses rigid behaviors such as forgetting prior corrections or reintroducing resolved errors. \ourmodel{} organizes memory around code-centric units and their iterative modifications, augmented with feedback from later rounds. \ourmodel{} combines the latest memory with similar past cases to form a compact session-level representation, and detects potential forgetting via AST-based change analysis, mitigating inconsistencies through LLM guidance.


Experiments on instruction following benchmark CodeIF-Bench for iterative code generation and the extended multi-turn repository-level code generation benchmark CoderEval demonstrate that \ourmodel{} improves current-turn and session-level instruction following by 12.2\% and 11.5\%, and can reduce interaction rounds about 2–3, significantly outperforming existing memory management and full context baselines. Moreover, \ourmodel{} has achieved highly competitive inference time and token cost.


The contributions are summarized as:
\begin{itemize}[leftmargin=*]
    \item We propose \ourmodel{}, a memory management system for repository-level iterative code generation that dynamically updates code context and mitigates forgetting during interaction.
    \item We introduce an AST-guided code memory strategy that preserves valid historical context and detects potential forgetting via code-change analysis.
    \item We conduct extensive evaluations on iterative code generation benchmarks, demonstrating state-of-the-art performance with competitive inference latency and token efficiency.
\end{itemize}

\section{Related Work}
\noindent \textbf{Repository-Level Code Generation.}
Code generation has evolved from standalone function synthesis~\cite{mbpp,humaneval} to repository-level settings~\cite{codereval,deveval}, where models exploit intra-repository context for coherent code generation. Prior work~\cite{repocoder,rl-coder,codeagent} has mainly focused on effective repository context retrieval. 
Although effective, these methods largely target single-round interactions. Recent benchmarks such as CodeIF-Bench~\cite{codeifbench} and SR-Eval~\cite{sreval} extend this setting to multi-round iterative scenarios, yet systematic approaches for improving LLM performance in such settings remain limited.

\noindent \textbf{Memory Systems for LLMs.}
Memory management has been proposed to support long-term interactions between LLMs and their environment~\cite{amem,mem0,memagent,memorybank}. Representative systems include MemGPT~\cite{memgpt}, which employs hierarchical memory, and MemoryOS~\cite{memos}, which introduces system-style multi-level storage and dynamic updates. However, these approaches are primarily designed for natural language dialogue. When applied to code generation, they often treat code as unstructured text, overlooking code structure, evolution, and code-centric interaction states, thereby limiting their ability to maintain consistency and stability in code changes.

\section{Methodology}
\begin{figure*}
	\centering
	\includegraphics[width=1\textwidth]{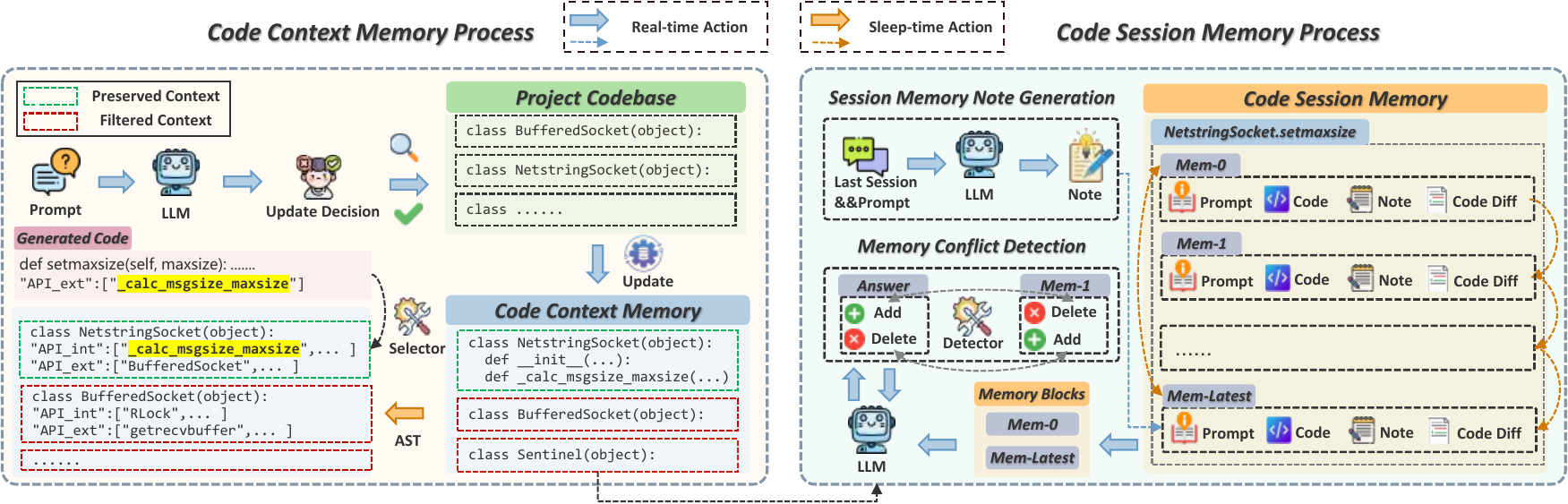}
 	\caption{The overall pipline of \ourmodel{}. }
	\label{fig:total_process}
    \vspace{-15pt}
\end{figure*}

\subsection{Task Definition}
In this section, we define the \textbf{repository-level iterative code generation} task. Given a code repository, an LLM incrementally generates and refines code through multi-turn interactions with the user. After each interaction round, the generated code is executed against a test suite to evaluate compliance with the current instruction.

Formally, given code repository $R$, user instruction $I$, and LLM-generated answer $A$, we get the following interaction sequence: $\{R,I_1,A_1,I_2,A_2,...I_n,A_n\}$. At round $i$, the correctness of $A_i$ is quantitatively assessed by executing the associated test suite $T_i$.

\subsection{Overview}
As shown in Figure~\ref{fig:total_process}, \ourmodel{} consists of two memory components: \textbf{Code Context Memory} for managing repository-level code context and \textbf{Code Session Memory} for managing session history. At the start of each session, \ourmodel{} uses the LLM to decide whether the code context requires updating and applies updates when necessary. Upon session completion, the \textbf{AST-based Selector} retains effective code contexts. Meanwhile, \ourmodel{} constructs session memory by combining the latest memory blocks with linked blocks based on prompt similarity. After code generation, the \textbf{AST-based Detector} identifies and mitigates potential memory conflicts, such as forgetting.

\subsection{Code Context Memory}
\subsubsection{Code Context Memory Block}

In this section, we describe the construction of the code context memory block. 
Using an AST parser\footnote{\url{https://docs.python.org/3/library/ast.html}}, \ourmodel{} decompose the repository into code blocks $m_{code}$ which corresponds to either a function or a class. 

Each block is represented as a key--value pair:
\begin{equation}
m_{\text{code}} =
\begin{cases}
\{( s_f ), v_f\}, & function\\[1mm]
\{(s_c, a_c, f_c), v_c\}, & class
\end{cases}
\end{equation}
where $s$ denote the function or class signatures with comments, $a_c$ denotes class attributes, $f_c$ denotes class function methods, and $v$ stores the complete implementation. \ourmodel{} utilizes keys when subsequently managing code context memory and employs values to generate responses.
This lightweight and efficient design minimizes input context length, improving decision quality while reducing inference overhead for downstream LLMs.

\subsubsection{AST-based Memory Selector}
The $Selector$ aims to preserve only effective historical contexts as memory. After round $t$, the $Selector$  first extracts external APIs from the generated code and collects internal and external APIs from memory entities. Let $\mathcal{A}(C_{t})_{\text{ext}}$ denote the set of external APIs for the generated code $C_{t}$, and let $\mathcal{M}_{\text{t}} = \{m_i\}$ be the code context memory, each associated with its APIs $\mathcal{A}(m_i)$ including both external and internal APIs. External APIs are methods that call others, while internal APIs can be called by others. 

The $Selector$ retains the $\mathcal{M}_{\text{t}}$ whose APIs intersect with the $C_{t}$'s external APIs:
\begin{equation}
\hat{\mathcal{M}_{\text{t}}} = \{\, m_i \in \mathcal{M}_{\text{t}} \mid \mathcal{A}(m_i) \cap \mathcal{A}(C_{t})_{\text{ext}} \neq \emptyset \,\}
\end{equation}
This ensures interface-level validity while preserving semantically related implementation patterns, enabling the LLM to effectively reuse relevant code structures in subsequent generation steps.

\subsubsection{Memory Processing}


At the beginning of round $t$, the LLM determines—using the prompt template in Figure~\ref{fig:code_context_judge}—whether the current code context memory $\mathcal{M}_{t-1}$ (initialized as empty) should be updated, based on $I$ including both historical and current instructions. 
This decision is made using only the \textit{key} representations $\{k_i\}$ extracted from the memory:
\begin{equation}
D = \text{LLM}(I, \{k_i\}) \in \{\texttt{ADD},\texttt{KEEP}\}
\end{equation}
where \texttt{ADD} indicates that a memory update is required and \texttt{KEEP} indicates that no update is necessary. If $D = \texttt{ADD}$, the relevant code blocks $\mathcal{M}_{\text{rel}}$ retrieved from the source code contexts according to the instructions are merged with the existing memory; if $D = \texttt{KEEP}$, the memory remains unchanged. The $\mathcal{M}_{t}$ can be denoted as:
\begin{equation}
\mathcal{M}_t =
\begin{cases}
\mathcal{M}_{t-1} \cup \mathcal{M}_{\text{rel}}, & \text{if } D = \texttt{ADD} \\[1mm]
\mathcal{M}_{t-1}, & \text{if } D = \texttt{KEEP} 
\end{cases}
\end{equation}

After interaction round $t$, the $Selector$  is applied to filter and preserve only valid context entities in $\mathcal{M}_{t}$ based on code $\mathcal{C}_t$, preventing redundant accumulation and improving generation stability and efficiency over long-term interactions.
\begin{equation}
\hat{\mathcal{M}_{t}} = Selector(\mathcal{M}_{t},\mathcal{C}_{t})
\end{equation}

\subsection{Code Session Memory}
\subsubsection{Session Memory Block}

We construct session memory blocks around code-centric units rather than natural language ~\cite{mem0,amem}. 
Given a conversation, its memory block is represented as:
\begin{equation}
m_{session} = \{ I, \mathcal{C}, \Delta \mathcal{C}, \mathcal{N}\}
\end{equation}
where $I$ is the user prompt, $\mathcal{C}$ is the generated function code, $\Delta \mathcal{C}$ is the code modification (diff) relative to the last round by AST-based analysis, and $\mathcal{N}$ is the note describing the modification generated by the LLM with the subsequent prompt. Feedback in the next round allows the LLM to generate more accurate notes, ensuring that unsatisfactory changes are properly corrected or correctness can be reused, thereby guiding subsequent edits. The block design also highlights the operational impact of prompts on code behavior and enables the LLM to track, reuse, and correct modifications in subsequent rounds.

In addition, iterative edits to the same function naturally link multiple memory blocks into a \textbf{memory sequence} ($ms_{\text{session}}$):
\begin{equation}
ms_{\text{session}} = \{ m_0, m_1, \dots, m_\text{latest} \}
\end{equation}
where $m_0$ correspondings to the initial function and subsequent blocks ordered by modification history and $m_\text{latest}$ correspondings to the latest memory.


\subsubsection{AST-based Memory Detector}

The $Detector$ aims to identify conflicts between newly generated code and historical session memory. The key idea is that inconsistencies between new code and stored memory blocks indicate memory conflicts and forgetting, exemplified by reverted corrections or reintroduced constraints.

First, instruction-level filtering is applied to exclude historical memory blocks corresponding to similar intents. 
Let $I_t$ denote the current instruction and $\{I_i\}_{i<t}$ denote historical instructions. 
We select the candidate detection session memory $\mathcal{M}_c$ such that:
\begin{equation}
\mathcal{M}_c = \{\, m_i \mid \text{sim}(I_t, I_i) < \tau \,\}
\end{equation}
where $\text{sim}(\cdot)$ denotes an instruction similarity function and $\tau$ is the filtering threshold. This filtering avoids spurious conflicts arising from iterative refinements or reversible changes under similar instructions.

Next, the $Detector$ extracts code changes $\Delta^t$ for current generated code $\mathcal{C}_{t}$ using AST-level diffs, where $\Delta^t = (\Delta^t_{\text{add}}, \Delta^t_{\text{del}})$
denote the sets of added and deleted AST nodes. Let $\Delta^i = (\Delta^i_{\text{add}}, \Delta^i_{\text{del}})$ denote the recorded code changes associated with the candidate detection memory $m_i \in \mathcal{M}_c$. 
The $Detector$ determines whether the session memory block $m_i$ is conflicted with the $\mathcal{C}_{t}$ by checking whether the current changes contradict its historical changes:
\begin{equation}
Conf(\Delta^t, \Delta^i) \triangleq
\left(
\Delta^t_{\text{del}} \cap \Delta^i_{\text{add}}
\;\;\cup\;\;
\Delta^t_{\text{add}} \cap \Delta^i_{\text{del}}
\right)
\end{equation}
where $Conf(\Delta^t, \Delta^i)$ captures AST nodes whose contradictory add–delete operations indicate forgetting or reversion of previously code changes.

Accordingly, the set of potentially conflicting session memory blocks is defined as:
\begin{equation}
\hat{\mathcal{M}_{c}} =
\{\, m_i \in \mathcal{M}_c \mid Conf(\Delta^t, \Delta^i)\neq \varnothing \,\}
\end{equation}

\subsubsection{Memory Processing}
At the end of round $t$, the interaction is recorded as a session memory block.
Let $\mathcal{C}_t$ denote the function code generated at round $t$, and let $MS=\{ms_i\}$ denote the set of existing memory sequences.
\ourmodel{} updates the session memory as:
\begin{equation}
MS \leftarrow
\begin{cases}
MS \cup \{\, m_t\,\}, & \mathcal{C}_t \notin MS \\
ms_{i} \oplus m_t, & \mathcal{C}_t \in ms_{i}
\end{cases}
\end{equation}
where $m_t$ stores the AST-level code diff $\Delta^t$ with respect to the latest version, and $\oplus$ denotes appending a block to an existing memory sequence. Within each memory sequence, blocks are linked by instruction similarity:
\begin{equation}
{Link}(m_t) = \{\, m_i \mid \text{sim}(I_t, I_i) \ge \tau \,\}
\end{equation}
This allows the latest block $m_t$ and its neighbors ${Link}(m_t)$ to serve as a compact yet information-rich representation.

At the beginning of round $t$, \ourmodel{} generates a modification note from $m_{t-1}$ conditioned on the current instruction and constructs the Code Session Memory as:
\begin{equation}
\mathcal{M}_{t} = \{ m_{t-1} \} \cup {Link}(m_{t-1})
\end{equation}
which is jointly used to guide code generation.
After a candidate solution $\mathcal{C}_{t+1}$ is produced, the $Detector$ identifies conflicting memory blocks:
\begin{equation}
\mathcal{M}_{\text{c}} =  Detector(I_{t+1},\mathcal{C}_{t+1})
\end{equation}
The final output $\hat{\mathcal{C}}_{t+1}$ is determined as:
\begin{equation}
\hat{\mathcal{C}}_{t+1} =
\begin{cases}
\text{LLM}(\mathcal{M}_{t},\mathcal{M}_{\text{c}}), & \text{if } \mathcal{M}_{\text{c}} \neq \varnothing \\[1mm]
\mathcal{C}_{t+1}, & \text{if } \mathcal{M}_{\text{c}} = \varnothing
\end{cases}
\end{equation}
where $\text{LLM}(\mathcal{M}_{t},\mathcal{M}_{\text{c}})$ represents that the LLM regenerates code based on $\mathcal{M}_{\text{c}}$.

\section{Experimental Setups}
In this section, we conduct experiments to evaluate the effectiveness of \ourmodel{}. We aim to answer the following research questions:
\begin{itemize}[leftmargin=*]
    \item \noindent\textbf{RQ1: Overall Performance.} How does \ourmodel{} perform overall in iterative repository-level code generation compared with baselines?
    \item \noindent\textbf{RQ2: Ablation Study.} How do the proposed memory components impact the performance of \ourmodel{}?
    \item \noindent\textbf{RQ3: Efficiency and Cost Analysis.} What are the time efficiency and computational cost of \ourmodel{}?
    \item \noindent\textbf{RQ4: Further Analysis for AST Parts.} The explainability Analysis of AST Components for \ourmodel{}.
\end{itemize}

\subsection{Benchmarks and Metrics}
\noindent\textbf{Benchmarks}. To evaluate \ourmodel{}'s effectiveness in repository-level iterative code generation tasks, we selected the iterative code generation instruction-following benchmark CodeIF-Bench~\cite{codeifbench} and the repository-level code generation benchmark CoderEval~\cite{codereval}. 

For CodeIF-Bench, we selecte the L-2 repository-level programming task, encompassing 40 dialogues with 360 instructions aligns with real-world development scenarios. Each dialogue centres around a python code generation task, comprising 9 distinct and non-conflicting instructions that can be tested and validated. CodeIF-Bench is to evaluate the LLM's ability to follow instructions during interactions: 1) The ability to follow the user's current instruction 2) The ability to follow session-level instructions 3) Errors arising from forgetting previously correct modifications during interaction (the phenomenon of forgetting).

For CoderEval, it comprises 230 python repository-level code generation tasks collected from real-world projects.
Following prior study~\cite{mint}, we extended CoderEval into multi-round iterative code generation tasks: we provide simple verbal feedback on code that fails testing to facilitate its re-generation. It simulates an iterative human–LLM interaction process in which a user repeatedly requests revisions to erroneous code, aiming to evaluate collaborative efficiency by measuring the number of user tasks completed within a fixed number of interaction rounds. Data examples are presented in the Appendix \ref{appendix:data_example}.

\noindent\textbf{Metrics}. We employ the following metrics~\cite{multiif,codeifbench} to evaluate both the correctness of LLM generated code as well as the LLMs' instruction following capability:
\begin{itemize}[leftmargin=*]
    \item \textbf{IA($\uparrow$)}: The LLM’s ability to follow the current instruction at each turn, measured by round-specific tests.
    \item \textbf{CA($\uparrow$)}: The LLM’s ability to satisfy all instructions issued so far, evaluated by cumulative tests.
    \item \textbf{IFR($\downarrow$)}: The proportion of previously satisfied instructions that fail in later turns, indicating forgetting.
    \item \textbf{Pass@k($\uparrow$)}: The probability that at least one of $k$ generated programs passes the tests in each iteration.
\end{itemize}

For further details, please refer to Appendix \ref{appendix:deatils of benchmarks and metrics}.

\begin{table*}[t]
\centering
\small
\setlength{\abovecaptionskip}{0.1cm}
\begin{tabular}{l|cccccccccc}
\toprule
Method & Turn-1 & Turn-2 & Turn-3 & Turn-4 & Turn-5 & Turn-6 & Turn-7 & Turn-8 & Turn-9 & Avg. \\
\midrule
\rowcolor[rgb]{ .741,  .843,  .933}
\multicolumn{11}{c}{Instruction Accuracy (IA)} \\ 
\midrule
 FC$_{\text{BM25}}$ & 37.5	& 40.0	&52.5	&45.0	&27.5	&\textbf{60.0}	&37.5	&35.0	&35.0   &41.1 \\
 FC$_{\text{RL-Coder}}$ & 35.0	& 37.5	&50.0	&37.5	&\textbf{35.0}	&60.0	&\textbf{47.5}	&35.0	&27.5	&40.6 \\
 MemGPT & 35.0	& 42.5	&42.5	& 42.5	& 27.5	& 50.0	&32.5	&40.0	& 40.0 & 39.1\\
 
 Mem0 & 37.5	& 27.5	& 37.5	& 32.5	& 20.0	 & 40.0	& 42.5	& 25.0	& 10.0	& 32.5\\
 A-Mem & 37.5	&42.5	& 45.0	& 47.5	& 27.5	 &50.0	& 30.0	& 47.5	& 42.5	& 41.1 \\
 \midrule
 \ourmodel{} & \textbf{40.0} &\textbf{50.0}	&\textbf{57.5}	&\textbf{50.0}	&32.5	&40.0	& \textbf{47.5}	&\textbf{50.0}	&\textbf{47.5}	&\textbf{46.1}\\
 w/o CtxMem  & 40.0  & 47.5  & 57.5  & 47.5  & 27.5  & 50.0  & 32.5  & 37.5  & 42.5  & 42.5\da{3.6} \\ 
w/o CtxAST & 40.0  & 42.5  & 47.5 & 47.5  & 30.0  & 50.0 & 37.5  & 40.0  & 40.0 & 41.7\da{4.4} \\ 
w/o SessAST & 40.0  & 42.5  & 62.5  & 47.5  & 27.5  & 42.5  & 47.5  & 47.5  & 40.0 & 44.2\da{1.9}\\ 
\midrule
\rowcolor[rgb]{ .741,  .843,  .933}
\multicolumn{11}{c}{Conversation Accuracy (CA)} \\ 
\midrule
 FC$_{\text{BM25}}$ & 37.5	& 38.8	& 43.3 &	43.1	& 35.5	 & \textbf{39.6} 	&37.9&	34.4 &	35.3 &	38.4 \\
 FCt$_{\text{RL-Coder}}$ & 35.0 & 33.8	 & 36.7	 & 36.9	& 34.5	& 36.2	& 37.5 & 	38.8 &  	 36.7 & 	36.2\\
 MemGPT & 35.0	& 36.3	&35.0	& 40.0	&33.5	&37.9	&34.3	&38.1	&36.4 &38.2\\
 Mem0 & 37.5	& 27.5	& 28.3	&28.7	&17.0	&16.7	&21.8	&25.3	&20.3	& 24.8\\
 A-Mem & 37.5	&41.3	&40.8	&43.8	&26.0	&38.8	&37.1	&36.6	&38.9	&37.8 \\
\midrule
\ourmodel{} & \textbf{40.0} & \textbf{47.5}	& \textbf{48.3} & \textbf{46.3}	& \textbf{37.0}	&37.9	& \textbf{41.1}  &\textbf{43.1}	&\textbf{44.4}	&\textbf{42.8}\\
w/o CtxMem  & 40.0  & 45.0  & 50.8  & 46.9  & 30.5  & 34.6  & 37.1  & 33.8  & 37.5 & 39.6\da{3.2}\\ 
w/o CtxAST & 40.0  & 42.5  & 41.7  & 39.4  & 36.5  & 36.3  & 37.7  & 37.2  & 37.4 & 38.7\da{4.1} \\
w/o SessAST & 40.0  & 43.8  & 50.8  & 46.3  & 33.5  & 36.2  & 39.3  & 42.2  & 40.8 & 41.4\da{1.4}\\ 
\bottomrule
\end{tabular}
\caption{The overall performance of LLMs in CodeIF-Bench. The metrics are IA and CA.}
\label{tab:RQ1_codeif}
\vspace{-0.3cm}
\end{table*}

\begin{table}[t]
\centering
\small
\setlength{\abovecaptionskip}{0.1cm}
\resizebox{\linewidth}{!}{
\begin{tabular}{l|ccccc}
\toprule
\multirow{1}{*}{Method}
 & Turn-1 & Turn-2 & Turn-3 & Turn-4 & Turn-5 \\
\midrule
FC$_{\text{BM25}}$ & 40.0	& 45.2	& 48.7 &	50.4	&50.9	 \\
FCt$_{\text{RL-Coder}}$ & 37.7	& 40.1	& 40.1 &	41.4	&41.4	 \\
A-Mem & 40.4 & 44.3	& 45.7 &	46.5	&46.5	 \\
Mem0 & 40.0	& 41.7	& 42.6 &	43.0	&43.0	 \\
MemGPT & 33.0	&42.2	& 46.5	&48.7	&49.1	\\

\midrule

\ourmodel{} & \textbf{42.2}	& \textbf{49.6}	& \textbf{51.7} & \textbf{54.3}	& \textbf{55.7}	 \\
w/o CtxMem & 42.2  & 47.8\da{1.8}   & 49.6\da{2.1}  & 50.0\da{4.3} &51.7\da{4.0}  \\
w/o CtxAST & 42.2  &  46.5\da{3.1}  & 50.0\da{1.7}  &  50.8\da{3.5} & 51.3\da{4.4}  \\ 
w/o SessAST & -  & - & - & - & -\\
\bottomrule
\end{tabular}
}
\caption{The overall performance of LLMs in CoderEval. The metric is Pass@1.}
\label{tab:RQ1_codereval}

\vspace{-0.3cm}
\end{table}

\subsection{Baselines and Implementation Details}
We selected the following baselines: Full-Context (FC), which uses all conversation history as context, and state-of-the-art memory management methods MemGPT~\cite{memgpt}, Mem0~\cite{mem0} and A-Mem~\cite{amem}. 
We also compared the single-turn code generation SOTA method RL-Coder, whose multi-turn configuration is identical to FC. We defined it as FCt$_{\text{RL-Coder}}$. The backbone model for all methods is DeepSeek-V3.2 with greedy decoding.
All methods except FCt$_{\text{RL-Coder}}$ use BM25 for code context retrieval. For memory retrieval and similarity, all memory-based methods—including \ourmodel{}—adopt the text-embedding-3-small vector retriever. The total number of documents for all retrieval settings is 5. The similarity threshold for \ourmodel{} is 0.95.

\section{Experimental Results}
\subsection{RQ1: Overall Performance}
\textbf{Performance on CodeIF-Bench.} Table~\ref{tab:RQ1_codeif} shows that \ourmodel{} consistently outperforms all baselines. For the IA score, it achieves the best performance in all rounds except 5 and 6. While memory management baselines such as MemGPT and A-Mem achieved results comparable to FC, their natural language–oriented memory representation limits their effectiveness, making them inferior to \ourmodel{}. Our advantage comes from dynamic code context updates and efficient session memory with accurate modification notes, enabling the LLM to better satisfy current instructions. Rounds 5 and 6 focus primarily on non-functional requirements, such as reducing circle complexity, which remain challenging for LLMs. The slight performance drop in these rounds indicates that \ourmodel{} integrates prior instructions effectively but struggles with complex non-functional constraints.
For the CA metric, \ourmodel{} outperforms all baselines except in round 6. Notably, while nearly all baseline methods, such as A-Mem and MemGPT, exhibit performance degradation as the number of interaction rounds increases, such as round 6, 7 and 8, \ourmodel{} demonstrates sustained improvement over time. This is due to the Code Session Memory mechanism that mitigates forgetting. Overall, these results highlight the effectiveness of \ourmodel{} in iterative repository-level code generation.

\textbf{Performance on CoderEval.} Table~\ref{tab:RQ1_codereval} presents results on the CoderEval benchmark. In the first round, performance differences largely stem from prompt design (Appendix~\ref{fig:meta_prompt}). In subsequent rounds, \ourmodel{} consistently outperforms all baselines due to dynamic code-context updating, which filters irrelevant code and integrates task-relevant context, and user-guided feedback, which generate more accurate reflections notes. These mechanisms reduce the number of interaction rounds needed to resolve bugs: for example, \ourmodel{} achieves the similar pass@1 score by round 3 that FC requires until round 5. Baselines like MemGPT and A-Mem, which performed similarly to FC on CodeIF-Bench, show weaker performance here, highlighting their limited generalizability. These results further confirm \ourmodel{}’s effectiveness for iterative, repository-level code generation.

\begin{SummaryBox}
\textbf{RQ1 Summary: }
\ourmodel{} consistently outperforms all baselines on both instruction following and iterative generation. Compared to the FC, \ourmodel{} improves instruction following ability by 12.2\% for IA and 11.5\% for CA and can alse reduce interaction rounds by 2–3.
\end{SummaryBox}

\subsection{RQ2: Ablation Study}
To evaluate the effectiveness of our AST-based memory components, we conduct an ablation study by systematically removing individual elements. Specifically, \textbf{CtxMem} denotes the full Code Context Memory, \textbf{CtxAST} the AST-based selector for Code Context Memory, and \textbf{SessAST} the AST-based detector for Code Session Memory.

On CodeIF-Bench (Table~\ref{tab:RQ1_codeif}), removing CtxMem—treating code context as static—substantially reduces performance, highlighting the importance of actively managing context during iterative generation. Removing CtxAST causes even larger degradation, as the LLM alone cannot filter irrelevant code and timely updates, leading to noisy context that misguides decision-making. Similarly, omitting SessAST consistently degrades performance, particularly in later rounds (e.g., CA in round~9 drops from 44.4\% to 40.8\%), due to increased forgetting of previously correctted changes.

For CoderEval (Table~\ref{tab:RQ1_codereval}), each session focuses on a single task and exhibits minimal forgetting; thus, SessAST ablations are unnecessary. Nevertheless, removing CtxMem or CtxAST still significantly impairs performance, whereas \ourmodel{} maintains consistent gains by dynamically preserving relevant code context. These results underscore the critical role of our AST-guided memory management in iterative repository-level code generation.

\begin{SummaryBox}
\textbf{RQ2 Summary: }
CtxMem greatly improves LLM performance by dynamically updating code context memory. Within CtxMem, CtxAST plays a crucial role by selectively preserving effective code context memory, while SessAST further enhances performance by identifying and mitigating memory forgetting.
\end{SummaryBox}
\subsection{RQ3: Efficiency and Cost Analysis}
\begin{table}[t]
\centering
\small
\setlength{\abovecaptionskip}{0.1cm}
\resizebox{\linewidth}{!}{
\begin{tabular}{lcccc}
\toprule
\multirow{2}{*}{\textbf{Method}} 
& \multicolumn{2}{c}{\textbf{CodeIF-Bench}} 
& \multicolumn{2}{c}{\textbf{CoderEval}} \\
\cmidrule(lr){2-3} \cmidrule(lr){4-5}
& \textbf{Avg. \#Time} & \textbf{Avg. \#Token}
& \textbf{Avg. \#Time} & \textbf{Avg. \#Token} \\
\midrule
FC$_{\text{BM25}}$        & 34.3 & 131.8k & 26.5 & 35.6k \\
FC$_{\text{RLCoder}}$    & 36.1& 72.5k  & 19.9& 19.5k \\
A-Mem             & 47.4 & 358.5k & 42.3 & 119.8k \\
Mem0             & 67.0 & 70.0k  & 55.2 & 8.0k \\
MemGPT             & 27.9 & 31.0k  & 38.5 & 61.3k \\
\midrule
\ourmodel{}         & 54.1 & 107.8k & 34.2 & 52.6k \\
w/o CtxMem       & 51.3 & 159.4k & 30.2 & 40.3k \\
w/o CtxAST       & 58.6 & 315.2k & 33.1 & 80.2k \\
w/o SessAST      & 36.4  & 75.1k  & -- & -- \\
\bottomrule
\end{tabular}
}
\caption{Average cost (in tokens) per data and completion time (in seconds) per round across methods.}
\label{tab:cost}
\end{table}

We further analyze the inference-time efficiency and token cost of \ourmodel{} (Table~\ref{tab:cost}). \ourmodel{} achieves competitive efficiency while delivering the best code generation quality. Baselines like A-Mem treat code context as natural language, incurring substantial inference overhead and higher token consumption. Mem0 compresses context to reduce tokens but requires many inference iterations, resulting in higher latency. MemGPT attains the lowest cost on CodeIF-Bench, yet its poor performance on CoderEval reflects unstable efficiency. FC with RL-Coder generally has lower time and cost than BM25 due to retrieval of shorter, finer-grained contexts.

\ourmodel{} leverages key–value–based code context representation, reducing token usage by ~30k on CodeIF-Bench compared to multi-turn dialogue methods. While slightly more costly than FC and Mem0 on CoderEval, it remains more efficient than other baselines. Ablation analysis shows that removing CtxMem slightly reduces inference time by eliminating LLM-controlled context management, but the key–value design adds only marginal overhead (2.8s per round on CodeIF-Bench, 4.0s on CoderEval) while substantially lowering token usage despite a partial increase on CoderEval. Omitting CtxAST increases time and cost further, even doubles token consumption on CodeIF-Bench, indicating that unfiltered context negatively affects LLM decision-making to continually introducing more irrelevant context.
Interestingly, removing SessAST substantially reduces cost and time, suggesting that iterative generation often produces conflicting code changes, which require additional validation. Exploring more efficient mechanisms to handle such contradictions is our future work.
\begin{SummaryBox}
\textbf{RQ3 Summary: }Compare to baselines, \ourmodel{} achieves competitive performance in terms of time efficiency and token cost. CtxMem introduces negligible time overhead and achieves net token savings despite occasional minor cost increases, and SessAST introduces additional cost and inference time to handle fogetting.
\end{SummaryBox}

\subsection{RQ4: Further Analysis for AST Parts}
\begin{figure}[t]
    \centering
    \begin{subfigure}{\linewidth}
        \centering
        \includegraphics[width=0.95\linewidth]{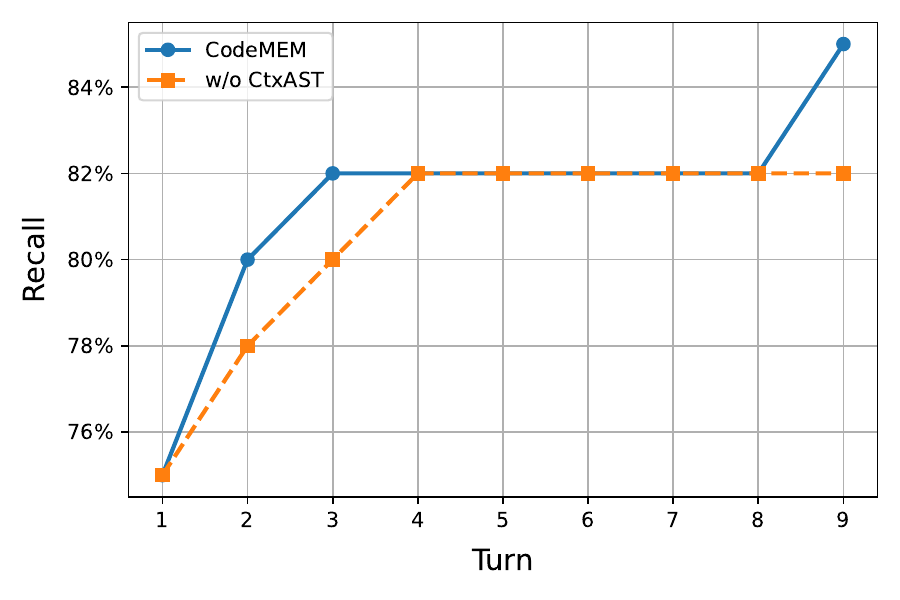}
        \caption{Recall}
        \label{fig:recall}
    \end{subfigure}

    \vspace{0.3em}  

    \begin{subfigure}{\linewidth}
        \centering
        \includegraphics[width=0.95\linewidth]{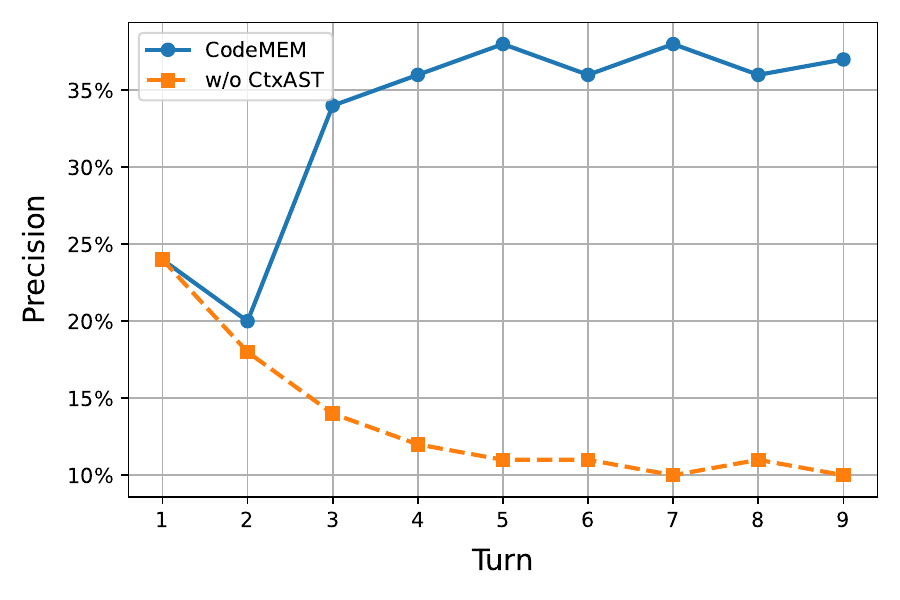}
        \caption{Precision}
        \label{fig:precision}
    \end{subfigure}

    \caption{Cumulative recall and precision per round for code contexts on CodeIF-Bench.}
    \label{fig:recall_precision}
\vspace{-0.4cm}
\end{figure}
\noindent\textbf{AST-based Memory Selector.} We evaluate the effectiveness of CtxAST by measuring recall and precision of relevant code in memory. On CodeIF-Bench, which provides ground-truth contexts, \ourmodel{} progressively retrieves more valid contexts while steadily improving precision (Figure~\ref{fig:recall_precision}). This improvement stems from the filtering effect of CtxAST. A temporary dip in precision in the second round is caused by noisy contexts during re-retrieval, but subsequent rounds show effective pruning. When CtxAST is removed and context is managed solely by the LLM, the LLM continuously retrieves and increases noise contexts, degrading generation quality and increasing both cost and inference time.

\begin{figure}[t]
    \centering
    \setlength{\abovecaptionskip}{0.1cm}
    \includegraphics[width=\linewidth]{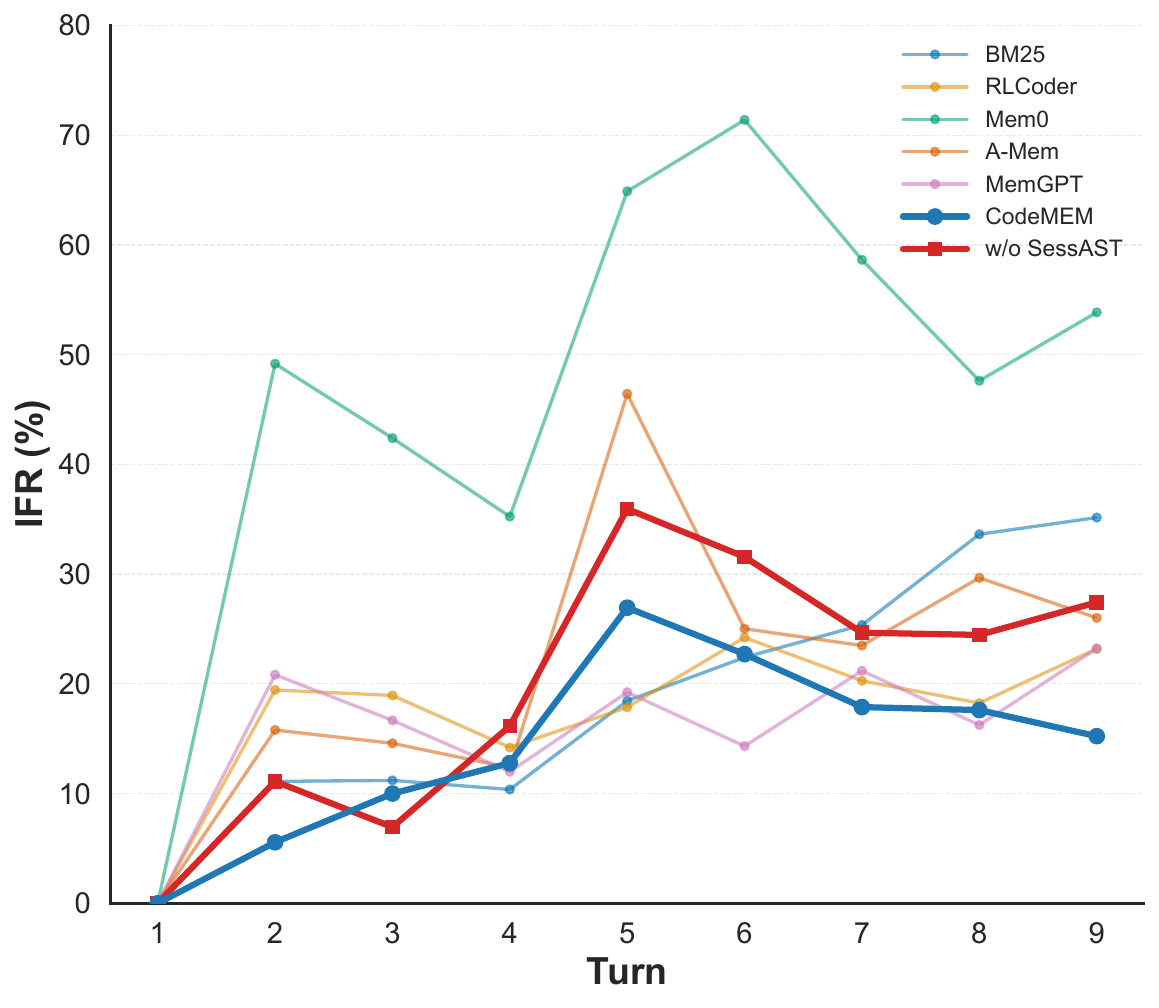}
	\caption{Forgetting rate (IFR) result on CodeIF-Bench.}
	\label{fig:ifr}
    \vspace{-0.5cm}
\end{figure}

\noindent\textbf{AST-based Memory Detector.} We assess SessAST using the forgetting rate (IFR) metric on CodeIF-Bench (Figure~\ref{fig:ifr}). \ourmodel{} substantially reduces forgetting compared to baselines, with improvements more pronounced in later rounds. Mem0 shows a sharp IFR increase over successive rounds, highlighting the limitations of natural language–based memory for iterative code tasks. Removing SessAST significantly increases forgetting, underscoring its critical role in code session memory. Moreover, \ourmodel{} demonstrates a sustained decline in IFR in later rounds, indicating its ability to mitigate forgetting in long-horizon, iterative code generation.

\begin{SummaryBox}
\textbf{RQ4 Summary: }
Further results demonstrate that CtxAST enhances the effective context recall and precision to deliver high-quality code context memory, while SessAST reduces the LLM's forgetting rate, thereby further improving the performance of \ourmodel{}.
\end{SummaryBox}

\section{Conclusion}
In this work, we propose \ourmodel{}, a memory management system tailored for repository-level iterative code generation. It dynamically preserves and incorporates effective code context through AST-guided Code Context Memory, and constructing code-centric Code Session Memory via AST to mitigate LLM forgetting. Experimental results on the CodeIF-Bench and CoderEval demonstrate that \ourmodel{} not only surpasses state-of-the-art baselines in quality but also achieves competitive time efficiency and cost consumption. In future work, we will further optimise \ourmodel{} to adapt it to more complex scenarios.

\section{Limitations}
\noindent\textbf{Limited Evaluation Scenarios.} Although our evaluation employs the instruction-following benchmark (CodeIF-Bench), which closely approximates real-world scenarios, and extends the existing code generation benchmark CoderEval in line with prior work, the following limitations remain: the efficacy of our model has not been validated across broader and more complex scenarios. For instance, longer interaction rounds and more intricate conversational scenarios. Constrained by the absence of such high-quality benchmarks, we shall continue to monitor and adapt \ourmodel{} to these benchmarks in future work.

\noindent\textbf{Biases Inherent to LLM.} \ourmodel{} partially relies on the LLM for decision-making and generating contextually relevant memories. This dependency on the LLM's inherent capabilities and preference limitations may lead to erroneous decisions and hallucinations. Furthermore, Code Context Memory relies on the LLM correctly utilizing context to select relevant information. The LLM may overlook pertinent context, thereby filtering out crucial elements. In future work, we will investigate methods to enhance the LLM's contextual utilization accuracy to mitigate this adverse effect.


\bibliography{ref}
\newpage
\appendix
\section{Metrics Details}
\label{appendix:deatils of benchmarks and metrics}
\begin{itemize}
    \item \textbf{Instruction Accuracy (IA)}~\cite{codeifbench}: This metric quantifies the proportion of instructions that the LLM correctly follows \textbf{in each round} of a dialogue. Specifically, in the $n$-th round, given the historical dialogue and the current instruction $I_n$, the LLM generates $A_n$. Compliance with instruction $I_n$ is determined by whether $A_n$ passes the corresponding test $T_n$: 
    \begin{equation}
        IA = 
        \begin{cases}
            1, & \text{if } A_n \text{ passes } T_n, \\
            0, & \text{otherwise.}
        \end{cases}
    \end{equation}
    \vspace{-5pt}
    
    \item \textbf{Conversation Accuracy (CA)}~\cite{codeifbench}: This metric measures the fraction of all instructions, from the first turn up to the current turn, that the LLM has successfully followed. In the $n$-th round, given the historical dialogue $\{I_1, A_1, \dots, I_{n-1}, A_{n-1}\}$ and the current instruction $I_n$, the LLM outputs $A_n$. The CA score for this turn is computed by evaluating $A_n$ against the full test sequence $TS = \{T_1, \dots, T_n\}$:
    \begin{equation}
        CA = \frac{\text{Number of tests in } TS \text{ passed by } A_n}{\text{Total number of tests in } TS}.
    \end{equation}

    \item \textbf{Instruction Forgetting Ratio (IFR)}~\cite{codeifbench}: This metric captures the proportion of previously followed instructions that the LLM fails to follow in later turns. An instruction is considered forgotten if it was satisfied in one of the previous rounds ($1, 2, \dots, n-1$) but not in the current round $n$. Let $PTS = \{T'_1, \dots, T'_k\}$ denote the tests passed in previous rounds. Then, given $A_n$, IFR is computed as:
    \begin{equation}
        IFR = \frac{\text{Number of tests in } PTS \text{ failed by } A_n}{\text{Total number of tests in } PTS}.
    \end{equation}

    \item \textbf{Pass@k}~\cite{humaneval}: This metric evaluates the LLM's performance by executing multiple generated programs per instruction:
    \[
        \text{Pass@}k = 1 - \frac{\binom{n - c}{k}}{\binom{n}{k}},
    \]
    where $k$ is the number of programs generated by the LLM for a given instruction, $c$ is the number of programs that pass the tests, and $n$ is the total number of programs generated.
\end{itemize}
 
\section{Data Example}
\label{appendix:data_example}

\begin{tcolorbox}[
breakable,
  colback=yellow!6!white,
  colframe=cyan!50!blue,
  boxrule=0.6pt,
  arc=2pt,
  left=6pt,
  right=6pt,
  top=6pt,
  width=\textwidth,
  bottom=6pt
]

\textbf{Example of CodeIF-Bench}

\vspace{0.4em}
\ttfamily\small
\{
\begin{adjustwidth}{2em}{0em}

\textbf{\textcolor{cyan!70!black}{namespace}}:
"boltons.socketutils.NetstringSocket.setmaxsize", \\[0.2em]
\textbf{\textcolor{cyan!70!black}{project\_path}}:
"Utilities/boltons", \\[0.2em]
\textbf{\textcolor{cyan!70!black}{completion\_path}}:
"Utilities/boltons/boltons/socketutils.py", \\[0.2em]
\textbf{\textcolor{cyan!70!black}{prompt}}: "Set the maximum size for receiving netstrings in the NetstringSocket instance. It updates the maxsize of the instance and calculates the maximum size for a netstring message based on the new maxsize value..."  \\[0.2em]

\textbf{\textcolor{cyan!70!black}{requirement}}: \{ \\[-0.2em]
\quad \textbf{Input-Output Conditions}: \{ \\[-0.2em]
\quad\quad \textbf{requirement}: "The 'setmaxsize' function should accept an integer 'maxsize' parameter and update the instance's 'maxsize' attribute accordingly...", \\[-0.2em]
\quad\quad  \textbf{test}: "tests/test\_socketutils.py::test\_setmaxsize\_updates\_attributes" \\[-0.2em]
\quad \}, \\[-0.2em]
\quad \textbf{Exception Handling}: \{ \\[-0.2em]
\quad\quad \textbf{requirement}: "The 'setmaxsize' function should raise a ValueError if the 'maxsize' parameter is not a positive integer or zero.", \\[-0.2em]
\quad\quad  \textbf{test}: "tests/test\_socketutils.py::test\_setmaxsize\_raises\_valueerror\_on\_invalid\_maxsize" \\[-0.2em]
\quad \}, \\[-0.2em]
\quad \textbf{Edge Case Handling}: \{ \\[-0.2em]
\quad\quad \textbf{requirement}: "The setmaxsize method should correctly handle setting the maximum size to zero and ensure that any non-empty netstring payloads cause a NetstringMessageTooLong exception.", \\[-0.2em]
\quad\quad  \textbf{test}: "tests/test\_socketutils.py::test\_setmaxsize\_zero\_behavior" \\[-0.2em]
\quad \}, \\[-0.2em]
\quad \textbf{Functionality Extension}: \{ \\[-0.2em]
\quad\quad \textbf{requirement}: "Extend the 'setmaxsize' function to print a message: 'Maxsize set to {new\_maxsize}' indicating the change in 'maxsize' for debugging purposes.", \\[-0.2em]
\quad\quad \textbf{test}test: "tests/test\_socketutils.py::test\_setmaxsize\_logs\_message" \\[-0.2em]
\quad \}, \\[-0.2em]
\quad \textbf{Annotation Coverage}: \{ \\[-0.2em]
\quad\quad \textbf{requirement}: "Ensure that the 'setmaxsize' function includes type annotations for its parameters and return type, including one parameters: 'maxsize': int, and return type: None.", \\[-0.2em]
\quad\quad  \textbf{test}: "tests/test\_socketutils.py::test\_setmaxsize\_annotations" \\[-0.2em]
\quad \}, \\[-0.2em]
\quad \textbf{Code Complexity}: \{ \\[-0.2em]
\quad\quad \textbf{requirement}: "The 'setmaxsize' function should maintain a cyclomatic complexity of 1, indicating a simple, linear function.", \\[-0.2em]
\quad\quad  \textbf{test}: "tests/test\_socketutils.py::test\_setmaxsize\_complexity" \\[-0.2em]
\quad \}, \\[-0.2em]
\quad \textbf{Code Standard}: \{ \\[-0.2em]
\quad\quad \textbf{requirement}: "The 'setmaxsize' function should adhere to PEP 8 standards, including proper indentation, line length, and spacing.", \\[-0.2em]
\quad\quad \textbf{test}: "tests/test\_socketutils.py::test\_check\_code\_style" \\[-0.2em]
\quad \}, \\[-0.2em]
\quad \textbf{Context Usage Verification}: \{ \\[-0.2em]
\quad\quad \textbf{requirement}: "The 'setmaxsize' function should utilize the '\_calc\_msgsize\_maxsize' method to update '\_msgsize\_maxsize'.", \\[-0.2em]
\quad\quad  \textbf{test}: "tests/test\_socketutils.py::test\_setmaxsize\_uses\_calc\_msgsize\_maxsize" \\[-0.2em]
\quad \}, \\[-0.2em]
\quad \textbf{Context Usage Correctness Verification}: \{ \\[-0.2em]
\quad\quad \textbf{requirement}: "Verify that the '\_msgsize\_maxsize' is correctly updated based on the new 'maxsize' using '\_calc\_msgsize\_maxsize'.", \\[-0.2em]
\quad\quad  \textbf{test}: "tests/test\_socketutils.py::test\_setmaxsize\_updates\_attributes" \\[-0.2em]
\quad \} \\[-0.2em]
\}
\end{adjustwidth}
\}
\end{tcolorbox}

\begin{tcolorbox}[
breakable,
  colback=yellow!6!white,
  colframe=cyan!50!blue,
  boxrule=0.6pt,
  arc=2pt,
  left=6pt,
  right=6pt,
  top=6pt,
  width=\textwidth,
  bottom=6pt
]

\textbf{Example of CoderEval in Our Experiment}

\vspace{0.4em}
\ttfamily\small
\{
\begin{adjustwidth}{2em}{0em}

\textbf{\textcolor{cyan!70!black}{\_id}}: "62e60f43d76274f8a4026e28", \\
\textbf{\textcolor{cyan!70!black}{file\_path}}: "neo4j/\_codec/hydration/v1/temporal.py", \\
\textbf{\textcolor{cyan!70!black}{project}}: "neo4j/neo4j-python-driver", \\
\textbf{\textcolor{cyan!70!black}{prompt}}: "Please finish the following code: def hydrate\_time(nanoseconds, tz=None): """for 'Time' and 'LocalTime' values..."  \\
\textbf{\textcolor{cyan!70!black}{feedback\_prompt}}: "Your answer is incorrect. Please regenerate."

\end{adjustwidth}
\}
\end{tcolorbox}
\clearpage
\section{Memory Example}
\begin{tcolorbox}[
breakable,
  colback=yellow!6!white,
  colframe=cyan!50!blue,
  boxrule=0.6pt,
  arc=2pt,
  left=6pt,
  right=6pt,
  top=6pt,
width=\textwidth,
  bottom=6pt
]

\textbf{Example of Code Context Memory}

\vspace{0.4em}
\ttfamily\small
\{
\begin{adjustwidth}{2em}{0em}

\textbf{\textcolor{cyan!70!black}{memory\_key}}: \{ \\[-0.2em]
\quad\textbf{class\_signature}: "class NetstringSocket(object): \\
\quad\quad Reads and writes using the netstring protocol \\
\quad\quad (see Wikipedia and protocol specification)...", \\[0.4em]
\quad \textbf{class\_attributes}: [
  "\_msgsize\_maxsize", "bsock", "maxsize", "timeout"
], \\[0.2em]
\quad \textbf{class\_methods}: [
  "\_\_init\_\_", "fileno", "settimeout", "read\_ns", "write\_ns"
] \\
\}, \\[0.6em]
\textbf{\textcolor{cyan!70!black}{memory\_value}}: "class NetstringSocket(object): \\
\quad def \_\_init\_\_(...): \\
\quad \quad self.bsock = BufferedSocket(...) \\
\quad \quad ... \\
\quad def read\_ns(...): \\
\quad \quad ..." \\
\end{adjustwidth}
\}
\end{tcolorbox}

\begin{tcolorbox}[
breakable,
  colback=yellow!6!white,
  colframe=cyan!50!blue,
  boxrule=0.6pt,
  arc=2pt,
  left=6pt,
  right=6pt,
  top=6pt,
width=\textwidth,
  bottom=6pt
]

\textbf{Example of Code Session Memory}

\vspace{0.4em}
\ttfamily\small
\{
\begin{adjustwidth}{2em}{0em}

\textbf{\textcolor{cyan!70!black}{boltons.socketutils.NetstringSocket.setmaxsize}}: [ \\[0.2em]
\quad \{ \\[-0.2em]
\quad\begin{adjustwidth}{1.5em}{0em}
\textbf{\textcolor{cyan!70!black}{id}}: 0, \\[0.3em]
\textbf{\textcolor{cyan!70!black}{instruction}}: "Please write a python function called 'setmaxsize' base the context...", \\[0.3em]
\textbf{\textcolor{cyan!70!black}{code}}: "def setmaxsize(self, maxsize): self.maxsize = maxsize    self.\_msgsize\_maxsize = self.\_calc\_msgsize\_maxsize(maxsize)" \\[0.3em]
\textbf{\textcolor{cyan!70!black}{note}}: "Function correctly implemented setmaxsize method.", \\[0.3em]
\textbf{\textcolor{cyan!70!black}{diff\_nodes}}: \{ \\[-0.2em]
\quad \textbf{\textcolor{cyan!70!black}{added}}: [\ ], \\[-0.2em]
\quad \textbf{\textcolor{cyan!70!black}{removed}}: [\ ] \\
\}, \\[0.4em]
\textbf{\textcolor{cyan!70!black}{state\_links}}: [\ ]
\quad\end{adjustwidth}
\quad \}, \\[0.2em]

...\\[0.2em]

\quad \{ \\[-0.2em]
\quad\begin{adjustwidth}{1.5em}{0em}
\textbf{\textcolor{cyan!70!black}{id}}: 2, \\[0.3em]
\textbf{\textcolor{cyan!70!black}{instruction}}: "The 'setmaxsize' function should raise a ValueError ....", \\[0.3em]
\textbf{\textcolor{cyan!70!black}{code}}: "def setmaxsize(self, maxsize): if not isinstance(maxsize, int) or maxsize < 0:        raise ValueError    self.maxsize = maxsize    self.\_msgsize\_maxsize = self.\_calc\_msgsize\_maxsize(maxsize)", \\[0.3em]
\textbf{\textcolor{cyan!70!black}{note}}: "The answer correctly implements the setmaxsize function as specified, raising a ValueError...", \\[0.3em]
\textbf{\textcolor{cyan!70!black}{diff\_nodes}}: \{ \\[-0.2em]
\quad \textbf{\textcolor{cyan!70!black}{added}}: [{'type': 'If+Raise', 'block': 'if not isinstance(maxsize, int) or maxsize < 0: raise ValueError("maxsize must be a non-negative integer")'}], \\[-0.2em]
\quad \textbf{\textcolor{cyan!70!black}{removed}}: [\ ] \\
\}, \\[0.4em]
\textbf{\textcolor{cyan!70!black}{state\_links}}: [\ ]
\quad\end{adjustwidth}
\quad \}, \\[0.6em]
\ ]  

\end{adjustwidth}
\}
\end{tcolorbox}

\clearpage
\section{Prompts}
\begin{tcolorbox}[
 title={The Code Context Memory Judge Prompt.},
  colback=yellow!6!white,
  colframe=cyan!50!blue,
  boxrule=0.6pt,
  arc=2pt,
  left=6pt,
  right=6pt,
  top=6pt,
  width=\textwidth,
  bottom=6pt
]

\textbf{You are an expert repository memory manager for repository code generation tasks.}

Your goal is to decide whether the current repository code context memory needs updating based on the user's programming instructions.

\vspace{0.5em}
\textbf{Decision Objective}

Decide if you need to modify the repository memory
(\textbf{Existing Repository Context}) based on how well it already covers
the entities mentioned in the user instructions.

\vspace{0.5em}
\textbf{Modes (Mutually Exclusive)}

\begin{itemize}
  \item \textbf{KEEP} — Use this mode when the existing repository context already contains
  all relevant classes/functions to understand or execute the instruction.
  
  \item \textbf{ADD} — Use this mode when Existing Repository Context lacks code context related to user instructions.
\end{itemize}

\textbf{User Instructions:}\\
\texttt{\{instructions\}}

\vspace{0.3em}
\textbf{Existing Repository Context:}\\
\texttt{\{existing\_repository\_context\}}



\vspace{0.5em}
\textbf{Output Format (strict JSON)}

\begin{tcolorbox}[
  colback=gray!5!white,
  colframe=gray!50,
  boxrule=0.4pt,
  arc=2pt
]
\ttfamily\small
\{\\
  "mode": "<ADD | KEEP>",\\
  "action": "<short, specific description of what to update or not update>",\\
  "target\_context": "<list of relevant namespaces or []>"\\
\}
\end{tcolorbox}
\label{fig:code_context_judge}
\end{tcolorbox}

\begin{tcolorbox}[
  title={The Genration Prompt with Memory.},
  colback=yellow!6!white,
  breakable,
  colframe=cyan!50!blue,
  boxrule=0.6pt,
  arc=2pt,
  left=6pt,
  right=6pt,
  width=\textwidth,
  top=6pt,
  bottom=6pt
]

\textbf{You are an expert repository-level code generator.}

Your goal is to generate the correct function implementation by leveraging
the provided repository context and historical memory blocks.

\vspace{0.6em}
\textbf{Repo Context}
\vspace{0.2em}

\begin{tcolorbox}[
  colback=gray!5!white,
  colframe=gray!50,
  boxrule=0.4pt,
  arc=2pt
]
\ttfamily\small
\{repo\_context\}
\end{tcolorbox}

\vspace{0.6em}
\textbf{Memory Blocks}
\vspace{0.2em}

\begin{tcolorbox}[
  colback=gray!5!white,
  colframe=gray!50,
  boxrule=0.4pt,
  arc=2pt
]
\ttfamily\small
\{memory\_blocks\}
\end{tcolorbox}

\vspace{0.6em}
\textbf{Current Instruction}
\vspace{0.2em}

\begin{tcolorbox}[
  colback=gray!5!white,
  colframe=gray!50,
  boxrule=0.4pt,
  arc=2pt
]
\ttfamily\small
\{instruction\}
\end{tcolorbox}

\vspace{0.6em}
\textbf{Output Requirement}

Please output the correct function implementation.
\label{fig:meta_prompt}

\end{tcolorbox}

\label{sec:appendix}

\end{document}